\documentclass[conference]{IEEEtran}
\bstctlcite{IEEEexample:BSTcontrol}
\ifCLASSINFOpdf

\else

\fi

\usepackage{graphicx}
\usepackage{amsmath,amssymb,epsfig}
\newcommand{\ds}{\displaystyle}

\begin{document}
	
	\title{q-LMF: Quantum Calculus-based Least Mean Fourth Algorithm}
    
\author{\IEEEauthorblockN{Alishba Sadiq\IEEEauthorrefmark{1}, Muhammad Usman\IEEEauthorrefmark{2}, Shujaat Khan\IEEEauthorrefmark{2}, Imran Naseem\IEEEauthorrefmark{1}\IEEEauthorrefmark{3}, \\ Muhammad Moinuddin\IEEEauthorrefmark{4} and Ubaid M. Al-Saggaf \IEEEauthorrefmark{5}}
		\IEEEauthorblockA{
%
			\\
			\IEEEauthorrefmark{1}College of Engineering, Karachi Institute of Economics and Technology, Korangi Creek, Karachi 75190, Pakistan.\\
			Email: \{alishba.sadiq, imrannaseem\}@pafkiet.edu.pk\\
			\\
			\IEEEauthorrefmark{2}Faculty of Engineering Science and Technology (FEST), Iqra University, Defence View, Karachi-75500, Pakistan.\\
			Email: \{musman, shujaat\}@iqra.edu.pk\\
			\\
			\IEEEauthorrefmark{3}School of Electrical, Electronic and Computer Engineering, The University of Western Australia, \\ 35 Stirling Highway, Crawley, Western Australia 6009, Australia.\\
			Email: imran.naseem@uwa.edu.au\\
		    \\
		    \IEEEauthorrefmark{4}Center of Excellence in Intelligent Engineering Systems (CEIES), King Abdulaziz University, Jeddah, Saudi Arabia.\\
		    Email: mmsansari@kau.edu.sa\\
		    \\
		    \IEEEauthorrefmark{5}Electrical and Computer Engineering Department, King Abdulaziz University, Jeddah, Saudi Arabia.\\ 
			Email: usaggaf@kau.edu.sa}}
%

	\maketitle
	
\begin{abstract}
	Channel estimation is an essential part of modern communication systems as it enhances the overall performance of the system.  In recent past a variety of adaptive learning methods have been designed to enhance the robustness and convergence speed of the learning process.  However, the need for an optimal technique is still there.  Herein,  for non-Gaussian noisy environment we propose a new class of stochastic gradient algorithm for channel identification.  The proposed $q$-least mean fourth ($q$-LMF) is an extension of least mean fourth (LMF) algorithm and it is based on the $q$-calculus which is also known as Jackson derivative.  The proposed algorithm utilizes a novel concept of error-correlation energy and normalization of signal to ensure high convergence rate, better stability and low steady-state error. Contrary to the conventional LMF, the proposed method has more freedom for large step-sizes. Extensive experiments show significant gain in the performance of the proposed $q$-LMF algorithm in comparison to the contemporary techniques.

\end{abstract}

	
	\IEEEpeerreviewmaketitle

\section{Introduction}\label{Sec:Intro}
The modern wireless communication systems provide a reasonable trade-off between performance parameters.  They provide high throughput with mobility while maintaining efficient utilization of limited bandwidth. Such cost-effective developments comes with a number of stumbling blocks and invigorating challenges.  Channel estimation is one such essential technique which is effectively used to enhance the performance of the modern wireless communication systems.  It is a widely used technique specifically in mobile wireless network systems as the wireless channel shows significant variations over time, generally these variations are caused by number of reasons such as transmitter or receiver being in motion at high speed.  Multipath interference from surroundings, such as highlands, buildings and other hindrances also affect the mobile wireless communication. In order to offer consistency, accuracy and high data rates at the receiver, accurate estimates of the time-varying channel is the requirement.  Linear models provide reasonable estimates with reduced Bit Error Rate (BER) thereby improving the capacity of the system \cite{FLMF}.  Adaptive learning methods are widely used to estimate the characteristics of the communication medium.  Due to their simplicity and ease of implementation the Least square-based methods are considered to be widely used optimization techniques for adaptive systems.  The technique has been applied in diversified applications such as function approximation \cite{khan2016novel}, detection of elastic inclusions \cite{W1}, noise cancellation \cite{NCLMS}, nonlinear system identification \cite{FBPTT}, ECG signal analysis \cite{ECGLMS}, elasticity imaging \cite{W3}, and time series prediction \cite{FRBF}, etc.  Adaptive filters are used to extract the desired components from a signal containing both desired and undesired components.  The least mean square (LMS) is a popular choice for designing adaptive filters. However, it has a slow convergence rate \cite{NLMS}.  
Numerous solutions has been designed to optimize the LMS algorithm \cite{BcLMS,LMS1,qLMS}.   In \cite{LMS11,FCLMS}, different solutions for complex signal processing were proposed.  Similarly, to deal with non-linear signal processing problem, the concept of kernel function-based LMS algorithms was proposed in \cite{KLMS1,TSPLMS,LMS4}.  

Beside these variants, various definitions of gradient have also been used to derive improved LMS algorithms; for instance in \cite{RVSSFLMS}, a robust variable step size fractional least mean square (RVSS-FLMS) based on fractional-order-calculus (FOC), is designed. The algorithm is derived using Riemann-Liouville fractional derivative for high convergence performance.  In \cite{VPFLMS,RVPFLMS}, some adaptive schemes were proposed for maintaining stability through adaptive variable fractional power.  The FOC variants are, however, not stable and diverge if the weights are negative or the input signal is complex \cite{bershad2017comments,CFLMScomments,wahab2018comments,mFLMScomments}.

Recently, a Jackson's derivative $q$-steepest descent algorithm is proposed that computes the normal to the cost function to achieve a higher convergence rate \cite{qLMS}.  The $q$-LMS algorithm has also been used for a number of applications, such as adaptive noise cancellation \cite{qLMS_ANC}, system identification, and designing of whitening filter \cite{qLMS_WF,qSSLMF}.\\  In \cite{qNLMS,EqLMS,Tv_qLMS}, adaptive frameworks are proposed for $q$ parameter.  In this research, we propose a modified variant of the stochastic gradient descent method by utilizing the $q$-steepest decent approach.  The proposed method is an extension of least mean fourth (LMF) algorithm which minimizes the fourth power of the instantaneous estimation error.  The conventional LMF can achieve higher convergence compared to the LMS algorithm \cite{LMF}.  However, it is inherently prone to instability due to the cubic power of the error signal in its update rule.  In the proposed $q$-calculus-based least mean fourth ($q$-LMF) algorithm the performance of the conventional LMF can be improved while maintaining the stability of the algorithm by using an additional controlling term $q$.
\subsection{Research Contributions and Paper Organization}
Following are the main contributions of this paper:
\begin{itemize}
\item[$\bullet$] A novel adaptive learning algorithm is derived for the identification of linear systems.  In particular a Jackson's derivative-based variant of LMF is derived using the $q$-gradient descent method \cite{qLMS}.
\item[$\bullet$] The reactivity of the proposed $q$-LMF algorithm is analyzed for $q$ controlling parameter. 
\item[$\bullet$] An interesting application of XE-NLMF filter is presented and a time-varying normalization technique is designed.
\item[$\bullet$] To study the transient and steady-state behaviors, the LMS and the LMF algorithms are compared to the $q$-LMF.
\item[$\bullet$] The  computational complexity of the proposed algorithm is also analyzed.  The $q$-LMF achieves significantly improved performance at the cost of very low computational overhead.
\item[$\bullet$] Performance claims are validated through computer simulations for a linear channel estimation task.
\end{itemize}

We organize the paper as follows.  A detailed overview of the $q$-calculus is explained in Section \ref{Sec:q-calculus}.  In Section \ref{Sec:Pro} the description of proposed algorithm is provided, followed by the experimental findings in Section \ref{Sec:Sim}.  The paper is concluded in Section \ref{Sec:Con}.

\section{Overview of \emph{q}-Calculus}\label{Sec:q-calculus}
The Quantum calculus is also referred to as the calculus without a limit \cite{qbook1}.  It has been successfully used in various areas including number theory, adaptive filtering, operational theory, mechanics, and the theory of relativity \cite{qbook2,qbook3,qbook4,qbook5}.

In q-calculus, the differential of a function is defined as (See, \cite{kac2001quantum})
\begin{equation}
	d_{q}(p(x)) = p(qx)-p(x).
\end{equation}
The derivative therefore takes the form
\begin{equation}
	D_{q}(p(x)) = \frac{d_{q}(p(x))}{d_{q}(x)} = \frac{p(qx)-p(x)}{(q-1)x}.
\end{equation}
When $q\rightarrow1$, the expression becomes the derivative in the classical sense.
For the form $x^{n}$ the q-derivative of a function is given as: 
\begin{equation}\label{q_derivative1}
	D_{q,x}x^{n} =\begin{cases}
		\ds\frac{q^{n}-1}{q-1} x^{n-1}, & q\neq1,
		\\ 
		\ds nx^{n-1}, & q = 1.
	\end{cases}
\end{equation}
The q-gradient of a function $p(x)$ for $n$ number of variables, $\mathbf{x} = [ x_{1}, x_{2},.... x_{n}]^{\intercal}$ is given as \cite{nthpower}
\begin{equation}\label{q_gradient1}
	\nabla_{q,w} p(x)\triangleq [D_{q1,x1}p(x),D_{q2,x2}p(x),...D_{qn,xn}p(x)]^{\intercal} ,
\end{equation}
where $q =[q_{1},q_{2},\dots q_{N}]^{\intercal}$.
	
	The q-calculus is considered to be a quite established field with well defined rules.  The product rule, for instance, takes two equivalent forms:
	\begin{multline}
	D_{q} (p(x)g(x)) = g(x)D_{q} p(x) + p(qx)D_{q} g(x) \\ = g(qx)D_{q} p(x) + p(x)D_{q} g(x). 
	\end{multline}
	Similarly, the quotient rule is given as:
	\begin{eqnarray}
	D_{q} \frac{p(x)}{g(x)} = \frac{g(x)D_{q} p(x) - p(x)D_{q} g(x)}{g(qx)g(x)},\quad g(x)g(qx)\neq 0. 
	\end{eqnarray}
	There is also a rule similar to the chain rule for ordinary derivatives. Let $g(x)=cx^{k}$.  Then
	\begin{eqnarray}\label{chainrule}
	D_{q} p(g(x)) = D_{q}^{k}(f)(g(x))D_{q}(g)(x).
	\end{eqnarray}
    There are various gradient descent based adaptive algorithms, $q$-calculus has been successfully used to develop $q$-gradient descent, outperforming the conventional gradient descent for real systems.

\section{Proposed \emph{q}-Least Mean Fourth (\emph{q}-LMF) Algorithm}\label{Sec:Pro}
By utilizing the idea of steepest descent with the following weight-update rule the conventional LMF algorithm is obtained
\begin{eqnarray}\label{LMS_WU}
\mathbf{w}(i+1) = \mathbf{w}(i) - \frac{\eta}{4} \nabla_{w}J(\mathbf{w}),
\end{eqnarray}
where $\eta$ is the step-size, $J(\mathbf{w})$ is the cost function for the $q$-LMF algorithm and is defined as
\begin{equation}\label{cost_function}
J(\mathbf{w})  = e^{4}(i),
\end{equation} 
$e(i)$ is the estimation error which is the deviation between the output signal at the $i^{th}$ instant and the desired response $d(i)$ i.e.,
\begin{eqnarray}
e(i) = d(i) - \mathbf{w}^{\intercal}(i) \mathbf{x}(i).
\end{eqnarray}
Here, $ \mathbf{x}(i)$ is the input signal vector defined as
\begin{equation}
 \mathbf{x}(i) = [x_{1}(i),x_{2}(i), \dots x_{M}(i)]^{\intercal},
\end{equation}
and $\mathbf{w}(i)$ is the vector consist of weights given as:
\begin{equation}
\mathbf{w}(i) = [w_{1}(i),w_{2}(i),\dots w_{M}(i)],
\end{equation}
where $M$ is the length of the filter.

 The $q$-LMF utilizes the Jackson derivative method \cite{qLMS}, it takes larger steps (for $q>1$) towards optimal solution. To derive $q$-LMF algorithm, conventional gradient in \eqref{LMS_WU} can be replaced with the $q$-gradient, we get
\begin{eqnarray}\label{q_gradient}
\mathbf{w}(i+1) = \mathbf{w}(i) - \frac{\eta}{4} \nabla_{q,w}J(\mathbf{w}).
\end{eqnarray}
 
The $q$-gradient of the cost function $J(\textbf{w})$ for the $k^{th}$ weight is defined as
\begin{eqnarray}\label{q_derivative}
	\nabla_{q,w_{k}}J(\mathbf{w}) = \frac{\partial_{q_k}}{\partial_{q_k} e}J(\mathbf{w}) \frac{\partial_{q_k}}{\partial_{q_k} y}e(i)\frac{\partial_{q_k}}{\partial_{q_k} w_{k}(i)}y(i).
\end{eqnarray}
Solving partial derivatives in \eqref{q_derivative} using the Jackson derivative defined in Section (\ref{Sec:q-calculus}) gives
\begin{multline}\label{first}
	\frac{\partial_{q_k}}{\partial_{q_k} e}J(w) = \frac{\partial_{q_k}}{\partial_{q_k} e}(e^{4}(i))\\= \frac{q_{k}^{4}-1}{q_{k}-1} e^{3}(i)= (q^{3}_{k}+q^{2}_{k}+q_{k}+1) e^{3}(i),
\end{multline}
where $J(\mathbf{w})  = e^{4}(i)$, we employ the instantaneous error term and it is given as
\begin{equation}
e(i) = d(i)-y(i)    
\end{equation}

Similarly
\begin{eqnarray}\label{second}
	\frac{\partial_{q_k}}{\partial_{q_k} w_{k}(i)}y(i) = \mathbf{x}_{k}(i),
\end{eqnarray}
and
\begin{eqnarray}\label{third}
	\frac{\partial_{q_k}}{\partial_{q_k} y}e(i) = -1,
\end{eqnarray}
Substituting equations \eqref{first}, \eqref{second}, and \eqref{third} in \eqref{q_derivative} gives
\begin{eqnarray}
	\nabla_{q,w_{k}}(i)J(w) = -(q^{3}_{k}+q^{2}_{k}+q_{k}+1)e^{3}(i)\mathbf{x}_{k}(i).
\end{eqnarray}
Similarly, for $k={1, 2, \dots, M}$,
\begin{multline}\label{q_derivative2}
	\nabla_{q,w} J(\mathbf{w}) = -(q^{3}_{1}+q^{2}_{1}+q_{1}+1)e^{3}(i)x_{1}(i),\\(q^{3}_{2}+q^{2}_{2}+q_{2}+1)e^{3}(i)x_{2}(i),  \\ \dots (q^{3}_{M}+q^{2}_{M}+q_{M}+1)e^{3}(i)x_{M}(i).
\end{multline}

Consequently, Eq. \eqref{q_derivative2} can be written as
\begin{eqnarray}\label{q_gradient2}
	\nabla_{q,w}J(w) = -4E[\mathbf{G} \mathbf{x}(i) e^{3}(i)],
\end{eqnarray}\label{diagonal_G}
where $\eta$ is the learning rate and $\mathbf{G}$ is a diagonal matrix
\begin{eqnarray}\label{G}
	{\rm diag}(\mathbf{G}) = [(\frac{q^{3}_{1}+q^{2}_{1}+q_{1}+1}{4}),\\ (\frac{q^{3}_{2}+q^{2}_{2}+q_{2}+1}{4}),.....(\frac{q^{3}_{M}+q^{2}_{M}+q_{M}+1}{4})]^{\intercal}.
\end{eqnarray}

Due to the ergodicity of the system \eqref{q_gradient2} results in
\begin{eqnarray}\label{ins_val}
	\nabla_{q,w}J(w) \approx -4G \mathbf{x}(i) e^{3}(i).
\end{eqnarray}
Substituting \eqref{ins_val} in \eqref{LMS_WU} renders the learning method of the $q$-LMF algorithm by
\begin{eqnarray}\label{qLMS_final}
	\boldsymbol{w}(i+1) = \mathbf{w}(i) +  \eta \mathbf{G} \mathbf{x}(i)e^{3}(i).
\end{eqnarray}

\subsection{Formulation of the $q$XE-LMF}
The least mean forth (LMF) algorithm shows better performance in non-Gaussian environment compared to the LMS \cite{LMF}.  For its stability normalized versions of the LMF algorithm are proposed that enhanced its performance and stability in Gaussian noisy environments. For example, the newly developed normalized LMF (XE-NLMF) algorithm is normalized by the error powers and mixed signal, and fixed mixed-power parameter is used to give it some weight \cite{XE-NLMF}.  As quantum calculus based algorithms shows faster convergence it is thought-provoking to expand the concept of normalization for the proposed $q$-LMF algorithm.\\
The proposed $q$XE-NLMF algorithm is expressed by the following equation:
\begin{eqnarray}\label{XE-NLMF}
\mathbf{w}(i+1) = \mathbf{w}(i) + \eta \mathbf{G} e^{3}(i) \mathbf{x}(i),
\end{eqnarray}
and 
\begin{equation}
{\rm diag}(\mathbf{G}) = 1/(\delta + \alpha \|x\|^{2} + (1-\alpha) \|e\|^{2}),
\end{equation}
where $\delta$ is a small value added to avoid indeterminate form, the notation $\|.\|^{2}$ shows the squared Euclidean norm of a vector and $\alpha$ is the mixing power parameter.
\subsection{Computational Complexity}
Computational complexity is an important performance measure of the learning algorithms.  We analyze the computation cost of the LMS, the LMF and the proposed $q$-LMF algorithm.  For each iteration the proposed $q$-LMF requires $3M+3$ multiplications and $2M$ additions. Which is $M+2$ and $M$ multiplications expensive compared to the LMS and LMF algorithms respectively.  Here $M$ denotes the number of unknown weight parameters.  The additional multiplications are required due to the presence of diagonal matrix $\mathbf{G}$.  Table \ref{table} shows the computational complexity of the LMS, the LMF, and the proposed $q$-LMF algorithm for one iteration.    Additionally, in initialization step the proposed $q$-LMF requires $3M$ multiplications and $3M$ additions only once to compute the $\mathbf{G}$ matrix.

\begin{table}[h]
\caption{Computational complexities of various algorithms}
\centering
\label{table}
\begin{tabular}{ccc}
\textbf{Algorithm} & \textbf{Multiplications} & \textbf{Additions} \\
LMS & 2M+1 & 2M \\
LMF & 2M+3 & 2M \\
Proposed q-LMF & 3M+3 & 2M
\end{tabular}
\end{table}

\section{Experiments}\label{Sec:Sim}
\begin{figure}[h]
		\centering
\centerline{\includegraphics[width=0.45\textwidth]{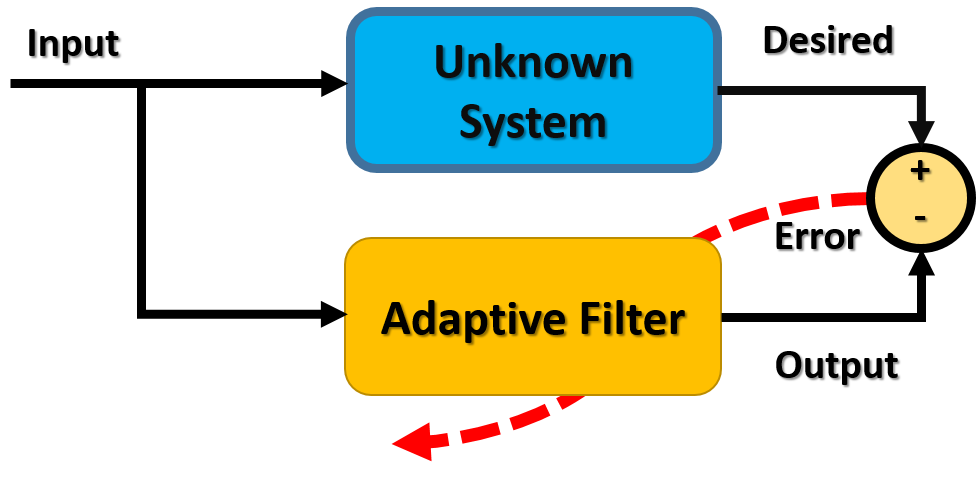} }
	\caption{adaptive channel estimation model.}
	\label{fig:Plant}
\end{figure}

For a system identification scenario, the performance of the proposed $q$-LMF algorithm is examined in this section.  Channel estimation, for instance, is a widely used method in communication systems to estimate the characteristics of an unknown channel.  A linear channel is displayed in Fig. \ref{fig:Plant}.
\begin{equation}\label{math_model}
y(t) = h_{1}x(t)+h_{2}x(t-1)+h_{3}x(t-2)+h_{4}x(t-3) + d(t).
\end{equation}
Equation \eqref{math_model} implies the mathematical model of the system, where the input and output of the system are denoted by $x(t)$ and $y(t)$  respectively and $d(t)$ is the disturbance which is taken to be white uniform noise in this case.  For this experiment, $\boldsymbol{x}(t)$ is chosen to be  $1\times10^{4}$ randomly generated samples obtained from Gaussian distribution of zero mean and variance of $1$.  For the simulation purpose the signal to noise ratio (SNR) is set to be 10 dB.  The experiments are repeated for $1000$ Monte Carlos independent runs and mean results are reported.  In equation \eqref{math_model}, the impulse response of the system is $h_{round\#}$ and in each simulation round random coefficients were assigned to the unknown channel and weights of the adaptive filter were reset to zero.

Following are the objectives of our simulations :
	\begin{description}
    	\item[$\bullet$] To show the performance gain of q-LMF over LMS in non-Gaussian environment. In particular, we compare the performance of proposed algorithm with the conventional LMS when $q$-LMF is operating as conventional LMF filter (i.e., $q=1$).
	\end{description}		\begin{description}
		\item[$\bullet$] To test the reactivity of the $q$-LMF algorithm over $q$ parameter.
	\end{description}
	\begin{description}
		\item[$\bullet$] To evaluate the performance of the proposed algorithm when $q$-LMF operating as XE-NLMF.
	\end{description}
For the performance analysis, the normalized weight difference (NWD) between the actual and the estimated weights is calculated. In particular, we define
\begin{equation}\label{NWD}
\mbox{NWD}=\frac{\left\Vert \bf h-\bf w \right\Vert}{\left\Vert\bf h\right\Vert},
\end{equation} 
where $\boldsymbol{w}$ is the obtained weight-vector and $\boldsymbol{h}$ is the actual impulse response of the channel.  
\subsection{The proposed $q$-LMF as the LMF algorithm}
\begin{figure}[h]
	\centering
	\centerline{\includegraphics[width=0.45\textwidth]{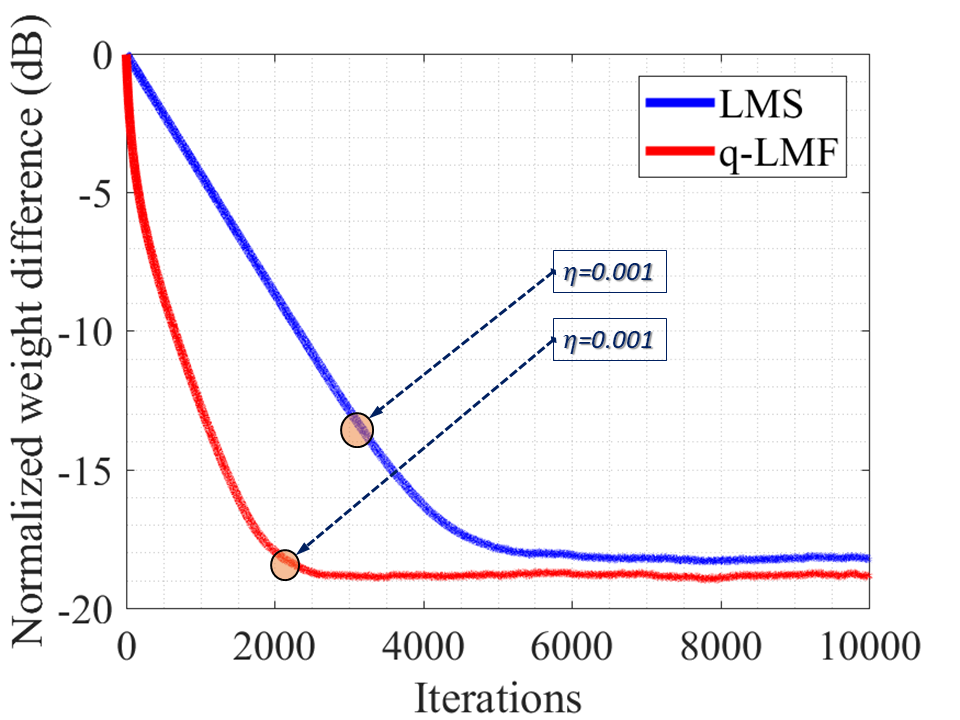} }
	\caption{Comparison of $q$-LMF with the traditional LMS algorithm.}
	\label{fig:LMS_qLMF}
\end{figure}

In this experiment, we compare the performance of $q$-LMF algorithm with the conventional LMS algorithm.  It is well established in \cite{LMF}, that when operating in a non-Gaussian environments the LMS perform inferior to the LMF.  For the evaluation of the proposed method in non-Gaussian environment, we choose the step-size $\eta=1e^{-3}$ for both the LMS and the proposed $q$-LMF with a fixed value of $q$ i.e. $1$. It can be seen from Fig.\ref{fig:LMS_qLMF} that when operating in non-Gaussian noise environment $q$-LMF also outperforms the LMS algorithm in steady-state error and convergence rate measures.  The proposed method converges at $2000^{th}$ iteration, whereas the convergence of the LMS algorithm is achieved at $4000^{th}$ iteration at a higher steady-state error compared to $q$-LMF. In steady-state error, the proposed $q$-LMF algorithm outperformed the LMS algorithm by a margin of approximately $0.6$ dB.  The steady-state error for LMS is $-18.239$ dB while for the proposed $q$-LMF it is relatively smaller i.e, $-18.827$ dB.  $q$-LMF exhibits low steady-state error and faster convergence rate in comparison to the traditional LMS (see table.\ref{prposedA}).
\begin{table}[!h]
\centering
\caption{Comparison of LMS and $q$-LMF with regard to Convergence point and steady-state error}
\label{prposedA}
\begin{tabular}{ccc}
\multicolumn{1}{l}{\textbf{Algorithms}} & \multicolumn{1}{l}{\textbf{Convergence point}} & \multicolumn{1}{l}{\textbf{Steady-state error (dB)}} \\
\textit{\textbf{LMS}}                   & 4000                                           & -18.239                                              \\
\textit{\textbf{q-LMF}}                 & 2000                                           & -18.827                                      
\end{tabular}
\end{table}

\subsection{Reactivity study of the proposed $q$-LMF algorithm}
\begin{figure}[h]
	\centering
	\centerline{\includegraphics[width=0.45\textwidth]{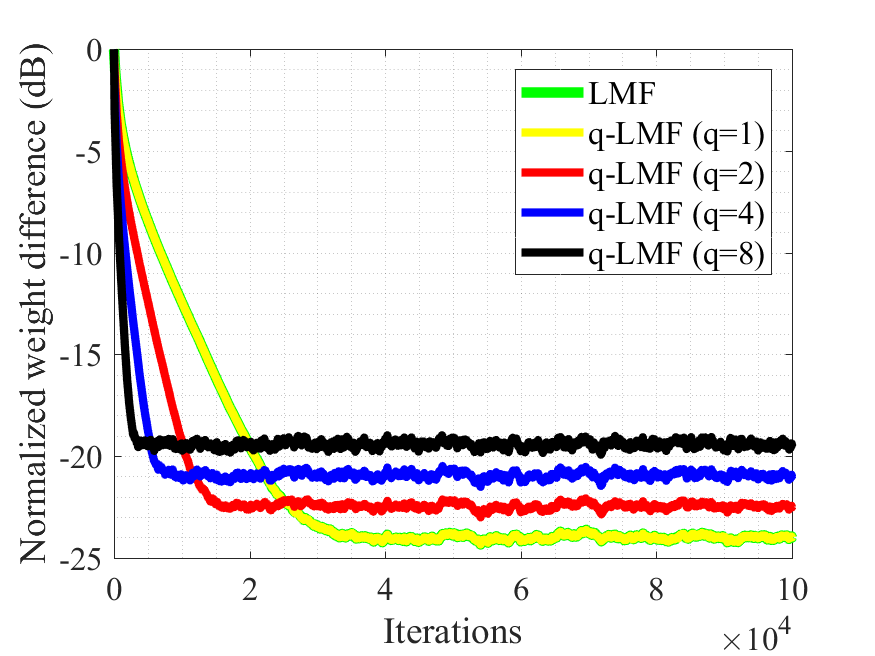} }
	\caption{NWD behaviour for the proposed $q$-LMF and the LMF algorithm.}
	\label{fig:reactivity}
\end{figure}

	 We observe the change in response of the $q$-LMF algorithm with the change in $q$ parameter.  Specifically, we conducted the simulations for a system identification problem and compare the normalized weight difference (NWD) learning graphs of the proposed $q$-LMF algorithm on various values of $q$ (see Fig.\ref{fig:reactivity}).  We considered four different values of $q$ i.e., $q=1$, $q=2$, $q=4$, and $q=8$. Fig \ref{fig:reactivity} shows that for $q=1$ the proposed $q$-LMF behaves exactly like the conventional LMF algorithm.  Note that for higher values of $q$, the proposed $q$-LMF shows faster convergence accompanied with a larger steady-state error. The $q$-LMF algorithm took the largest number of iterations for $q=1$ i.e, at $28000$ while for greater values of $q$ it took lesser iterations such as for $q=2$, $q=4$ and $q=8$ it took $14000, 6000$ and $2500$ iterations to converge.(refer table \ref{point}, )

\begin{table}[!h]
\centering
\caption{Convergence point for different values of $q$}
\label{point}
\begin{tabular}{cc}
\multicolumn{1}{l}{\textbf{Value of $q$}} & \multicolumn{1}{l}{\textbf{Convergence point (number of iterations)}} \\
$q$ = 1                                   & 28000                                                                 \\
$q$ = 2                                   & 14000                                                                 \\
$q$ = 4                                   & 6000                                                                  \\
$q$ =  8                                  & 2500                                                                 
\end{tabular}
\end{table}
	
\subsection{The proposed $q$-LMF as XE-LMF}
\begin{figure}[h]
	\centering
	\centerline\fbox{{\includegraphics[width=0.45\textwidth]{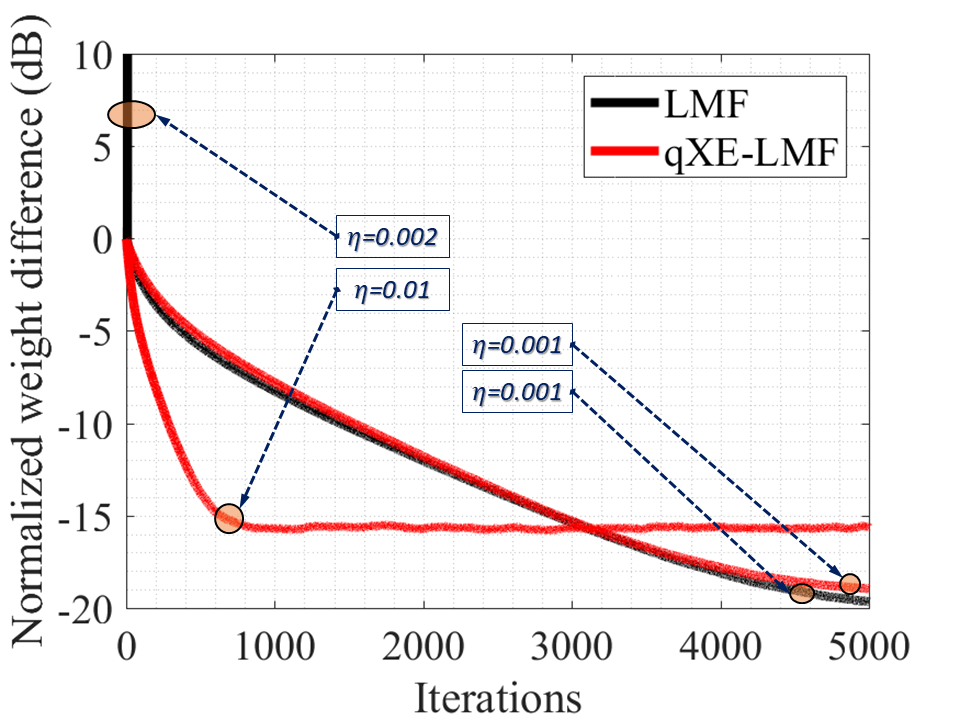} }}
	\caption{Comparison of the conventional LMF and the proposed $q$XE-LMF.}
	\label{fig:LMF_qLMF2}
\end{figure}

Stability of the LMF algorithm is difficult to achieve,  since the cubic power of the error ($e^{3}$) in the LMF gradient vector can create overwhelming initial uncertainty.  To solve this problem we propose to use the $q$-LMF in XE-NLMF mode, this will help $q$-LMF to operate at higher values of step size on which the conventional LMF diverges and the proposed algorithm can achieve better performance.  In this section, we show the comparison of the LMF and the proposed $q$-LMF algorithm for large and small values of step-size.  In Fig. \ref{fig:LMF_qLMF2}, when operating at $\eta$ = $0.001$ it can be noticed that both the algorithms show slow convergence .  For the performance evaluation of the proposed algorithm at a higher convergence rate, we simulated the LMF algorithm for same simulation setup with $2$ times greater value of step-size i.e., $\eta$ = $0.002$ at which it shows the divergence while the proposed $q$-LMF when operating in XE-NLMF mode (derived in \eqref{XE-NLMF}) can operate at even higher convergence speed (i.e., $\eta$ = $0.01$).  Overall,  the proposed $q$-LMF algorithm when operating in XE-NLMF mode can achieve higher convergence while maintaining the stability.

\section{Conclusion}\label{Sec:Con}
In this research, we proposed a $q$-calculus based LMF algorithm called $q$-LMF.  The proposed algorithm provides additional control over the convergence and steady-state performances through $q$ parameter.  The standard LMS and LMF algorithms are compared to the proposed $q$-LMF   for a problem of channel estimation in non-Gaussian environment.  The algorithms are compared on the basis of steady-state error, convergence rate and computational complexity.  The simulations were repeated for $1000$ independent Monte Carlos simulation rounds at $10$ dB SNR value.  Overall, the proposed $q$-LMF algorithm comprehensively outperformed the LMS, and the LMF algorithms, achieving the best steady-state error and convergence rate performances.
	
	
	\bibliographystyle{IEEEtran}
	\bibliography{References}

\end{document}